\documentclass[12pt]{iopart}
\usepackage{cite}
\usepackage{epsf}  
\begin{document}

\title[superconductivity and the  spin correlations in  Na$_{x}$CoO$_{2}$$\cdot 1.3$H$_{2}$O]
{Na content dependence of  superconductivity and the  spin correlations in  Na$_{x}$CoO$_{2}$$\cdot 1.3$H$_{2}$O}

\author{
Guo-qing Zheng$^1$,
Kazuaki Matano$^1$,\\
R L Meng$^2$,
J Cmaidalka$^2$ and
C W Chu$^{2, 3, 4}$
}

\address{
$^1$ Department of Physics, Okayama University, Okayama 700-8530, Japan}

\address{
$^2$ Department of Physics and TCSAM,
University of Houston,\\
TX 77204-5932, USA}

\address{
$^3$ Lawrence Berkeley National Laboratory,
1 Cyclotron Road, Berkeley,\\
CA 94720, USA
}

\address{
$^4$ Hong Kong University of Science and Technology, Hong Kong, China
}

\ead{\mailto{zheng@psun.phys.okayama-u.ac.jp}}

\begin{abstract}\\
We report systematic measurements using $^{59}$Co nuclear quadrupole resonance (NQR) technique  on the  cobalt oxide superconductors Na$_{x}$CoO$_{2}$$\cdot 1.3$H$_{2}$O over a wide Na content range $x$=0.25$\sim$ 0.34. We find that $T_c$ increases with decreasing $x$ but reaches to a plateau for $x \leq$0.28. In the sample with $x \sim$0.26, the spin-lattice relaxation rate $1/T_1$ shows a $T^3$ variation below $T_c$ and down to $T\sim T_c/6$, which unambiguously indicates  the presence of  line nodes in the superconducting (SC) gap function. However,  for  larger or smaller $x$, $1/T_1$ deviates from the $T^3$ variation below $T\sim$ 2 K even though the $T_c$ ($\sim$4.7 K) is similar, which suggests an unusual evolution of the SC state.  
In the normal state,  the  spin correlations   at a finite wave vector  become stronger upon decreasing $x$, and the density of states at the Fermi level increases with decreasing $x$, which can be understood in terms of a single-orbital picture suggested on the basis of LDA calculation.

\end{abstract}




\section{Introduction}
The layered cobalt oxides Na$_{x}$CoO$_{2}$  have attracted much attention in recent years, because of a rich variety of orders competing for the ground state. For $x$=0.75, the material shows a large thermoelectric power \cite{Terasaki,Wang}, while
the $x$=0.5 compound is a charge ordered insulator with $T_{CO}$=50 K \cite{Foo}.
The magnetic properties are different for different Na contents \cite{Foo,Barnhardt,Boothroyd}.  At the side with   $x$ larger than 0.5, the DC susceptibility shows a Curie-Weiss temperature ($T$) variation, while at the low- $x$ side    the susceptibility is $T$-independent \cite{Foo}. 
Most interestingly, superconductivity with $T_c\sim$4.5 K emerges around $x$=0.3 when water molecules are intercalated in between the Co layers \cite{Takada}. It has  been reported that $T_c$ can be tuned by varying $x$ in the range of $0.25\leq x \leq$0.38   \cite{Shaak}. 

One of the unsettled issues on the superconductivity is the symmetry of the gap function. Theoretically, a chiral($d+id$) superconducting state was first proposed on the basis of RVB scenario \cite{Baskaran,WangLeeLee,Ogata}. 
In a previous study, we found that the spin-lattice relaxation rate $1/T_1$ shows no coherence peak below $T_c$, and follows a $T^n$ variation with $n$= 2.2$\sim$1 at low $T$ \cite{Fujimoto}.   This suggests that the superconductivity is of non-BCS type, but it has been pointed out that the data are  insufficient to distinguish between some exotic states \cite{Baratsky,ZDWang,Singh0}.  It has been shown that  both chiral  and pure $d$-wave states in the presence of  disorder \cite{Baratsky,ZDWang}, or an odd-frequency $s$-wave state \cite{Singh0} can be compatible with such  $T$-variation of $T_1$.  Since identification of the gap function is the first step toward the understanding of the superconductivity,  more work is needed.
 
Another issue is how the electron correlations in the normal state evolve with doping, which should help understanding the mechanism for the superconductivity. 
In the octahedral crystal electric field environment, two $e_g$  orbitals of Co are located much higher in energy above  $t_{2g}$ orbitals.  The local density approximation (LDA) band calculation shows that \cite{Singh,Zou,Pickett}, among the three $t_{2g}$ orbitals, the $a_{1g}$ ($3d_{3z^2-r^2}$, where $z$-axis is the threefold rotation axis) orbit lies   higher in energy than the remaining two $e^,_g$  orbitals. If one takes this picture \cite{Baskaran,WangLeeLee}, then the Co$^{4+}$ state without Na has one electron  in the $a_{1g}$ orbital (spin 1/2),  with the $e^,_g$ orbitals fully occupied. This single-orbital situation resembles the undoped cuprates;  adding Na dopes electrons into the $a_{1g}$ orbital. 
In contrast, multiple-band models were also proposed \cite{KKM,Yanase}, which draw similarities to the manganese oxides, another important class of strongly correlated materials.
 
In this letter, we address the above two issues by studying the evolution of the system with doping using NQR (nuclear quadrupole resonance) technique. We present intensive  data on  the superconducting  and the normal states over a wide range of   Na content.

\section{Experimental Results and Discussion}
The Na$_x$CoO$_2$$\cdot$yH$_2$O powder \cite{Lorenz} was synthesized following Ref. \cite{Takada}. 
The precursor of Na$_{0.75}$CoO$_{2}$  was  immersed in the Br$_2$/CH$_3$CN solution to de-intercalate Na. The amount of Br$_2$ is listed in Table 1, where X means the theoretical value necessary to remove all  the Na. The Na content of the final product was determined by the ion coupled plasma method.
 X-ray diffraction 
 indicates that the samples are single phase  except the 5X ($x\sim$0.34) and 10X ($x\sim$0.32) samples in which minor secondary phases are present. The c-axis elongates as the Na content decreases, in agreement with previous report \cite{Shaak}.
$T_c$ was determined from the ac susceptibility and the $1/T_1$ measurements. 
The previously reported sample \cite{Fujimoto} was made with 10X Br amount in a separate  run and turned out to have an $x\sim$0.31. All other samples were made in the same run.
$^{59}$Co NQR measurements were carried out by using a phase-coherent spectrometer. 
 The nuclear magnetization decay curve is excellently fitted to the theoretical curves \cite{Mac}, with a unique $T_1$ component.

\begin{table}
\caption{\label{Table 1} Characterizations of the samples.  }
\begin{indented}
\lineup
\item[]\begin{tabular}{@{}*{7}{l}}
\br

Br amount & Na content  $x$ & c-axis length (\AA) & a-axis length (\AA) & $T_c$ (K) \\ \hline
5X  & 0.34 & 19.50 & 2.822 & 2.65 \\ \hline
10X & 0.32 & 19.62 & 2.822 & 3.60 \\ \hline
10X \cite{Fujimoto} & 0.31 & 19.59 & 2.819 & 3.70 \\ \hline
20X & 0.28 & 19.73 & 2.817 & 4.75 \\ \hline
40X & 0.25 & 19.80 & 2.817 & 4.70 \\ \hline
100X & 0.26 & 19.79 & 2.815 & 4.60 \\ 
\br
\end{tabular}
\end{indented}
\end{table}

Figure 1(a) shows a typical example of the three NQR transition lines arising from the nuclear spin $I$=7/2 of $^{59}$Co. Figure 1(b) shows the $\pm$ 3/2$\leftrightarrow$$\pm$5/2 transition line for various Na contents. 
 Firstly,  the peak frequency increases as $x$ decreases down to $x$=0.28. Analysis of three transition lines allows us to find that  the  $\nu_Q$ increases as the Na content decreases down to $x$=0.28, but the asymmetry parameter $\eta$ decreases monotonically with decreasing Na content, as seen in Fig. 1(c) and (d), respectively. Here $\nu_{Q}$ and $\eta$ are defined as $\nu_{Q}\equiv\nu_{z}=\frac{3}{2I(2I-1)h}e^2Q\frac{\partial ^2V}{\partial z^2}$,  $\eta=|\nu_{x}-\nu_{y}|/\nu_{z}$,  with   $Q$ being the nuclear quadrupole moment and $\frac{\partial ^2V}{\partial \alpha^2} (\alpha=x, y, z)$ being the electric field gradient  at the position of the nucleus \cite{Abragam}. 
 The decrease of $\nu_Q$ with increasing $x$ is consistent with doping electrons into the $a_{1g}$ orbit \cite{Matano-0}.
 Secondly, for the 5X ($x\sim$0.34) and 10X ($x\sim$0.32) samples, there appear multiple peaks for the transition. However, the additional peaks with smaller amplitude are not due to the minor impurity phase detected in the X-ray chart. Rather, we find that $1/T_1$ measured in  different peaks  differs only by 10\% in magnitude and shows the identical temperature variation; a clear decreases was seen below $T_c$. These  peaks may have arisen from the ordering of Na ions \cite{Zandbergen} 
 which are located above and below Co. More detailed discussion will be published elsewhere \cite{Matano-0}.
 We note in passing that the multiple peaks seen in the $\pm$ 3/2$\leftrightarrow$$\pm$5/2 transition are not resolvable  in the $\pm$ 5/2$\leftrightarrow$$\pm$7/2 transition.

\begin{figure}
\begin{center}
\epsfxsize=10cm
\epsfbox{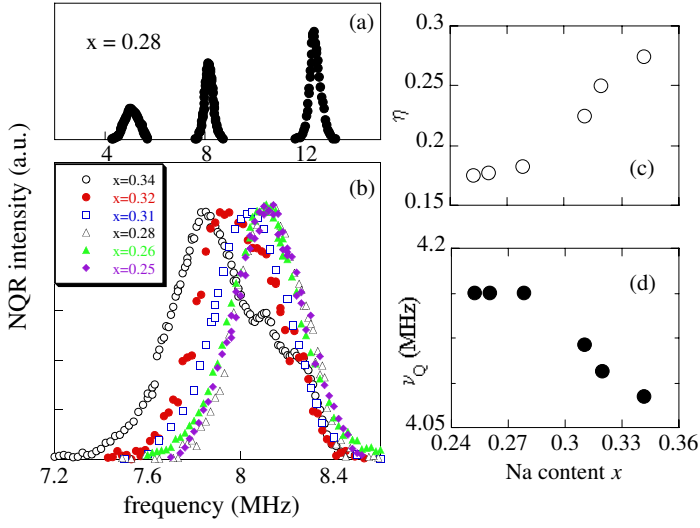}
\caption{\label{Fig.1} (a) $^{59}$Co  NQR spectra for the sample of $x\sim$0.28. (b) The line shape of the $\pm 3/2 \leftrightarrow \pm$5/2 transition for various Na contents. (c) and (d) Na content dependencies of   the asymmetric parameter $\eta$ and $\nu_Q$, respectively.}

\end{center}
\end{figure}

Figure 2 shows the temperature dependence of the quantity $1/T_1T$ for various $x$. All measurements were done at the $\pm$ 3/2$\leftrightarrow$$\pm$5/2 transition. For the 5X and 10X samples, $1/T_1$ was measured at the main peak shown in Fig. 1(b). Two trends can be seen in Fig. 2. First, the absolute value increases with decreasing $x$. Second,  $1/T_1T$ increases upon lowering $T$,  indicating the importance of spin correlations;    the smaller the  $x$ the  steeper the increase, indicating that the spin correlations become stronger as $x$ is reduced. 
 
 What is the nature of the spin correlations?
One good  way to tell  is through  inspection of the Knight shift, which measures the spin susceptibility and, unlike the bulk susceptibility,   is not affected by the presence of magnetic impurities. 
In a superconducting single crystal with $x\sim$0.3, we found that  the shift  is $T$-invariant from $T_c$ up to $T$=120 K \cite{Matano}, which is in good agreement with the report of Ning {\it et al}  \cite{Ning}. These results rule out the possibility of ferromagnetic spin fluctuations with wave vector $q$=0, since in that case the Knight shift would follow the temperature dependence of  $1/T_1T$ because both of them would be dominated by the  susceptibility at wave vector $q \sim$ 0. 

\begin{figure}
\begin{center}
\epsfxsize=10cm
\epsfbox{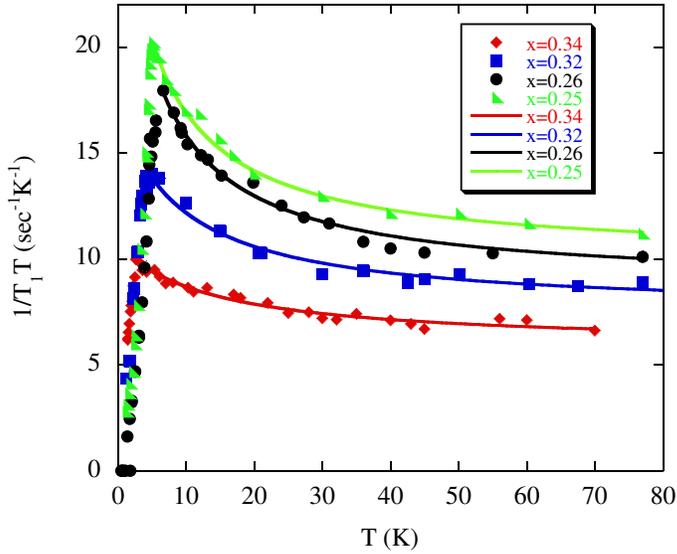}
\caption{\label{Fig.2} The quantity $1/T_1T$ as a function of temperature for various Na contents.   The curves are fits to $1/T_1T = (1/T_1T)_0+ C/(T+\theta)$, with resulting parameter $C$=75, 77, 111 and 123 Sec$^{-1}$ for $x$=0.34, 0.32, 0.26 and 0.25, respectively. The values of $(1/T_1T)_0$ and $\theta$ are plotted in Fig. 3. }

\end{center}
\end{figure}

The results thus indicate that the spin fluctuations  peak at finite wave vector. The most likely candidate is  antiferromagnetic (AF) correlations \cite{Baskaran,WangLeeLee,Ogata,NS}, although its detailed $q$-dependence awaits clarification by future experiments, in particular neutron scattering. 
Below we analyze our data in terms of AF correlations, in an attempt to quantify the discussion. In a general form, $1/T_1T$ is written as $1/T_1T=\sum_q A_q^2 \chi^{,,} (q,\omega)/\omega$ in the $\omega \rightarrow$0 limit, where $A_q$ is the hyperfine coupling constant, and $\chi^{,,} (q,\omega)$ is the imaginary part of the $q$-dependent, dynamical susceptibility. 
If one assumes that there is a peak around the AF wave vector $Q$, then one may have the following approximation: $1/T_1T =(1/T_1T)_{AF}+(1/T_1T)_{0}$. Here $(1/T_1T)_0$ denotes the contribution from wave vectors other than $Q$, which usually is dominated by the uniform susceptibility $\chi_0$. For the AF contribution, we use the formula for a two-dimensional nearly antiferromagnet, namely, ($1/T_1T)_{AF}=\frac{C}{T+\theta}$ (Ref. \cite{Moriya}). We find that such model fits the data quite successfully, as depicted by the curves in Fig. 2. The resulting fitting parameters are shown in Figure 3. Note that a small value of $\theta$ means that a system is close to the AF instability ($\theta$=0). The enhancement of the  spin correlation (decrease of $\theta$) with decreasing $x$, as is the case in doped copper oxide high-$T_c$ superconductors,  can be naturally understood in the single-orbital picture \cite{Baskaran,WangLeeLee} described earlier. In  contrast, the multiple-band theory predicts Na content-insensitive, ferromagnetic spin correlations \cite{Yanase}.

Turning to the quantity $(1/T_1T)_0$, its increase with decreasing $x$ can be attributed to the increase of density of state (DOS) at the Fermi level. This result is consistent with the LDA band calculation which found that the dominant Fermi surface is a {\it hole-like} one and that the DOS  increases   as the electron number is reduced \cite{Zou,Zhang}. This aspect is different from the case of doped cuprates where the DOS decreases as decreasing carrier density. 
We summarize our results in Fig. 3. 
 As $x$ decreases from 0.34, $T_c$ increases, but reaches to a plateau of $\sim$4.7 K for  $x\leq$0.28. Our results for high value of $x$   are in good agreement with the report  by Schaak {\it et al} \cite{Shaak}.
 A plateau has also been reported recently by Chen {\it et al} and Milne {\it et al} \cite{Chen,Milne}.  It is interesting to note that as the parameter $\theta$ decreases and $C$ increases, namely, as the electron correlation becomes stronger, $T_c$ increases.
 
 \begin{figure}
\begin{center}
\epsfxsize=8cm
\epsfbox{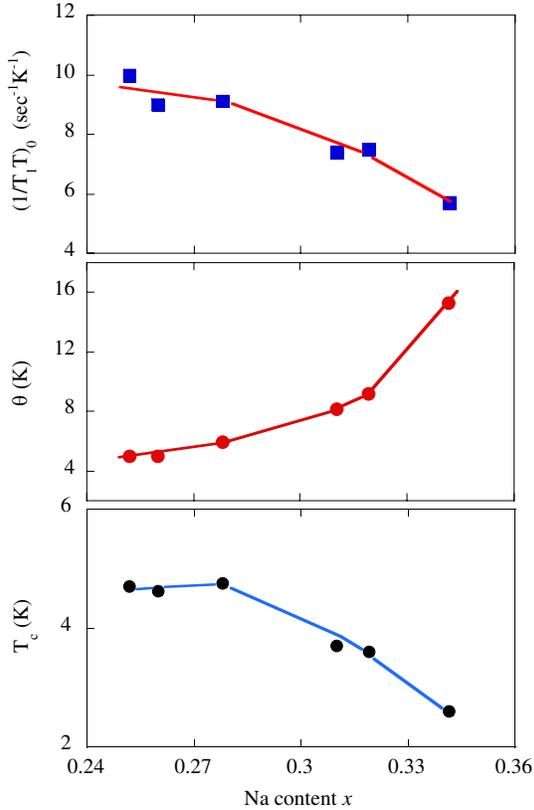}
\caption{\label{Fig.3} Various parameters obtained in this study plotted as a function of Na content, $x$. The curves are guides to the eyes.}
\end{center}
\end{figure}

Next, we turn to the superconducting state. 
As can be seen already in Fig. 2, no coherence peak was seen just below $T_c$ in all samples, irrespective of Na content and/or $T_c$ value. However, the low-$T$ behavior is sample dependent.  We first show in Fig. 4 the results for
  the sample with $x \sim$0.26. It can be seen that $1/T_1$ follows a $T^3$ variation down to the lowest temperature measured ($\sim T_c/6$). This result indicates unambiguously that there exists line-nodes in the gap function.
As we noted in  previous publications \cite{Fujimoto,Zheng}, the gap function with line nodes generates    an energy ($E$)-linear DOS at low $E$ which results in a $T^n$ ($n$=3) dependence of $1/T_1$.  In the previous sample, $1/T_1$  did not follow an exact $T^3$ variation  and became proportional to $T$ below $T\sim T_c/2$,  \cite{Fujimoto}. It has been proposed that many exotic superconducting states can be compatible with the data \cite{Baratsky,ZDWang,Singh0}. Calculations incorporating   impurity/disorder scattering  showed that  a $d_{x^2-y^2}$+i$d_{xy}$ state which has a full gap is more consistent with the previous data \cite{Baratsky}. 
 Our present result completely rules out the  $d_{x^2-y^2}$+i$d_{xy}$ state proposed by the $t-J$ model in an isotropic triangular lattice \cite{Baskaran,WangLeeLee,Ogata, Feng}.  Also, the odd frequency  state proposed \cite{Singh0} is  gapless \cite{Mazin,Fuseya}, and is  incompatible with our finding in the $x$=0.26 sample. 
However, if anisotropy of the $t$ term is incorporated, the theories  could be compatible  with our experimental results. In fact, it has been recently shown that a small anisotropy in $t$ (10$\sim$20\%)  lifts the degeneracy between $d_{x^2-y^2}$ and $d_{xy}$ states, leading to a stabilization of a pure $d$-wave state \cite{Watanabe}. 

\begin{figure}
\begin{center}
\epsfxsize=10cm
\epsfbox{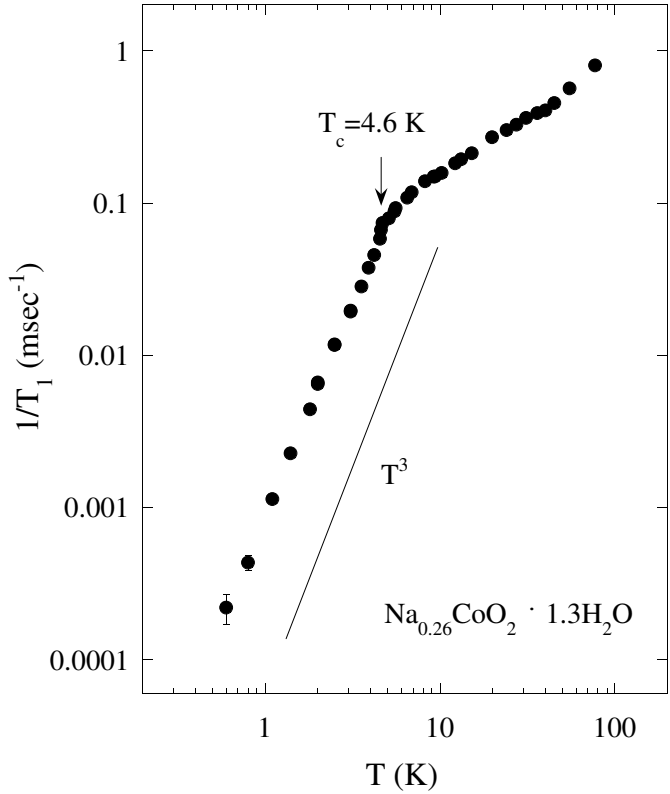}
\caption{\label{Fig.4} Temperature dependence of $1/T_1$ for the sample of  $x$=0.26. The arrow indicates $T_c$. The straight line depicts the $T^3$ variation.}
\end{center}
\end{figure}

\begin{figure}
\begin{center}
\epsfxsize=10cm
\epsfbox{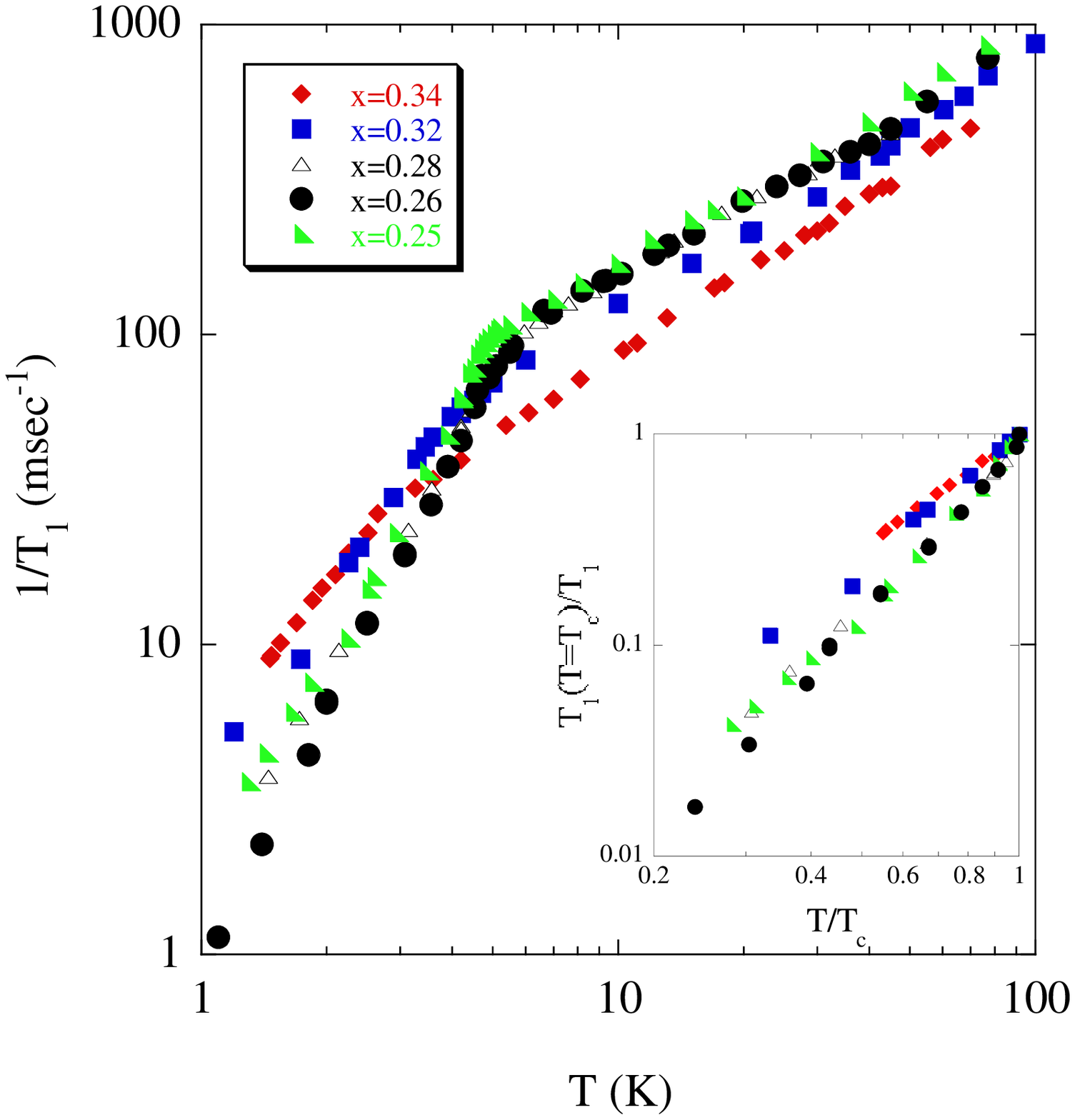}
\caption{\label{Fig.5} A comparison of the temperature dependence of $1/T_1$ in the superconducting state for various samples with different Na content. The inset shows the normalized $1/T_1$ by its value at $T=T_c$ against the reduced temperature.}
\label{Fig.5}
\end{center}
\end{figure}
 In other samples  with either larger or smaller $x$, however, $1/T_1$ deviates from the $T^3$ relation below $T$=2.0 K.  To see this more clearly, we plot the $T$-dependence of $1/T_1$ for all samples in 
Fig. 5. In the inset of the figure is shown the normalized $1/T_1$ by its value at $T=T_c$ against the reduced temperature.  
What is responsible for the deviation from the $T^3$ variation in samples with $x$ either larger or smaller than 0.26? Disorder or impurity is an unlikely candidate, since the samples with $x\sim$0.28 and $x\sim$0.25 have very similar or even higher $T_c$ and NQR line-width as the $x \sim$ 0.26 sample. 
Clearly, this unprecedented property deserves further investigation, both experimentally and theoretically. Meanwhile,
low-lying excitations due to nearby competing orders may be a possible candidate. For larger $x$, it is known that charge ordering \cite{Foo,Zandbergen} occurs. 
The Na order could also give rise to the anisotropy of the hopping integral $t$ mentioned earlier.
  For smaller $x$, on the other hand, 
it has  been proposed that different charge/magnetic order may occur at $x$=1/4 \cite{Baskaran2}. 

\section{Conclusion}

In conclusion, we have presented intensive  $^{59}$Co- NQR measurements on the  cobalt oxide superconductors Na$_{x}$CoO$_{2}$$\cdot 1.3$H$_{2}$O over a wide Na content range $x$=0.25$\sim$ 0.34. We find that $T_c$ increases with decreasing $x$ but reaches to a plateau for $x \leq$0.28. In the superconducting state, the spin-lattice relaxation rate $1/T_1$ shows a $T^3$ variation down to $T\sim T_c/6$  for $x \sim$0.26, which indicates unambiguously the presence of  line nodes in the superconducting gap function. For both larger or smaller $x$, however, $1/T_1$ deviates from the $T^3$ dependence even through the $T_c$ value is similar. This unprecedented finding shows the richness of the physics of this new class of superconductors, and any pertinent theory must explain this peculiar property.  In the normal state, the temperature dependence  of $1/T_1T$ indicates that the  spin correlations at finite wave-vector become stronger upon decreasing $x$, while the DOS increases. This aspect is different from the case of doped cuprates but can be understood in terms of  a single-orbital picture suggested on the basis of LDA calculation.

\ack
We thank Y. Sun for x-ray chart analysis, G. Baskaran, Y. Fuseya, H. Harima, Y. Kitaoka, G.Khaliullin, C.T. Lin, K. Miyake,  M. Ogata and Y. Yanase  for  helpful discussions. This work was supported in part by research grants from MEXT, NSF, the T. L. L. Temple Foundation, the John and Rebecca Moores Endowment, and DoE.

\end{document}